\documentclass{article}
\usepackage{spconf,amsmath,graphicx}

\usepackage{array}
\newcolumntype{P}[1]{>{\centering\arraybackslash}p{#1}}

\usepackage{multirow}
\usepackage{booktabs}
\usepackage{amssymb}

\usepackage{enumitem}
\setlist{topsep=0pt, leftmargin=*}
\usepackage{colortbl,hhline}
\usepackage{xcolor}
\usepackage{authblk}
\usepackage{hyperref}
\usepackage{float}
\usepackage{color,soul}

\title{Pre-training strategies using contrastive learning and playlist information for music classification and similarity}

\name{Pablo Alonso-Jim\'enez$^{1, 2}$,  Xavier Favory$^{1}$, Hadrien Foroughmand$^{1}$, Grigoris Bourdalas$^{1}$, Xavier Serra$^{2}$, Thomas Lidy$^{1}$, Dmitry Bogdanov$^{2}$}

\address{
    Utopia Music, Switzerland$^{1}$\\
    Music Technology Group, Universitat Pompeu Fabra, Spain$^{2}$
    }

\begin{document}
%
\maketitle

\begin{abstract}
In this work, we investigate an approach that relies on contrastive learning and music metadata as a weak source of supervision to train music representation models.
Recent studies show that contrastive learning can be used with editorial metadata (e.g., artist or album name) to learn audio representations that are useful for different classification tasks.
In this paper, we extend this idea to using playlist data as a source of music similarity information
and investigate three approaches to generate anchor and positive track pairs.
We evaluate these approaches by fine-tuning the pre-trained models for music multi-label classification tasks (genre, mood, and instrument tagging) and music similarity.
We find that creating anchor and positive track pairs by relying on co-occurrences in playlists provides better music similarity and competitive classification results compared to choosing tracks from the same artist as in previous works.
Additionally, our best pre-training approach based on playlists provides superior classification performance for most datasets.

\end{abstract}

\section{Introduction}
\label{sec:intro}

Learning better representations is crucial to improve the quality of music classification and similarity models.
Many popular approaches apply end-to-end models to learn representations while optimizing classification objectives~\cite{dieleman2014end, choi2016automatic}. Other directions include pre-training models on editorial metadata~\cite{Park2018RepresentationLO,Kim2018,Lee2019,Kim2020,huang2020large}, multi-modal correspondence~\cite{cramer2019look}, co-listening statistics~\cite{huang2020large}, contrastive supervised~\cite{Favory2020COALACA,favory21coala, Ferraro2021Enriched} and self-supervised~\cite{spijkervet2021contrastive,niizumi2021byol,yao2022contrastive,zhao2022s3t,wang2022towards} objectives, music generative models~\cite{castellon2021codified}, playlist co-occurrences~\cite{Ferraro2021Enriched}, text~\cite{manco2021learning}, or combinations of them~\cite{huang2020large,Kim2020,castellon2021codified,Ferraro2021Enriched}.
Recently, contrastive learning has shown promising results in audio and music representation learning, especially in self-supervised fashions~\cite{yao2022contrastive,zhao2022s3t}, and some studies suggest that it allows learning more robust features than classification objectives~\cite{wang2022towards}.

Scientific evidence suggests that, in contrastive setups, it is beneficial to choose positive pairs that share information relevant for the downstream task while being diverse with respect to irrelevant characteristics~\cite{tian2020makes}.
However, most audio and music self-supervised contrastive methods rely on sample mixing~\cite{wang2022towards}, audio effects~\cite{spijkervet2021contrastive}, or temporal crops~\cite{Saeed2021Constrastive} to generate the augmented versions, which intuitively have a small potential to obtain samples that are distinct enough. 

Accounting for this observation, a recent study inspired by COLA~\cite{Saeed2021Constrastive} shows that selecting the positive pairs according to editorial metadata co-occurrences (e.g., songs from the same artist) improves the learned representations significantly~\cite{alonso2022music}.
In this work, we extend this method to operate with new sources of music metadata.
Specifically, we focus on music consumption metadata in the form of playlists.
We propose strategies to obtain positive pairs by
(i) randomly sampling tracks co-occurring in playlists, (ii) constraining the positive pairs to the top co-occurrences across playlists, and (iii) using alternative track representations obtained using a Word2Vec~\cite{mikolov2013efficient} model trained on the playlist sequences as associated pair.
We pre-train models based on the ResNet50~\cite{he2016deep} and VGGish~\cite{hershey2016cnn} architectures with playlists from the Million Playlist Dataset~\cite{chen2018recsys} (MPD) and then transfer the learned representations to solve music classification and similarity tasks.


The main contributions of this work are the following:
\begin{itemize}[noitemsep]
    \item 
    We compare the performance of three models based on playlist data and four baselines using two different architectures in one similarity and five classification tasks.
    \item 
    We propose pre-training strategies using playlist information that lead to superior performance compared to previous approaches based on editorial metadata in several music classification tasks.
    \item 
    We show that some models trained with playlists achieve better similarity metrics than those based on self-supervision or editorial metadata.
\end{itemize}
 
The rest of this manuscript is organized as follows: Section~\ref{sec:motivation} provides further motivation for the exploration of consumption metadata as a source of supervision, Section~\ref{sec:method} describes the proposed pre-training methods, Section~\ref{sec:experiments} provides details about the experimental setup, and in Section~\ref{sec:results} we present and discuss the results. Finally, Section~\ref{sec:conclusion} outlines the principal conclusions of this work.
 
\vspace{-0.7em}
\section{Motivation}
\label{sec:motivation}
\vspace{-0.7em}


Self-supervised approaches enable training models with a large amount of unannotated data, which has been successful in fields such as natural language processing~\cite{qiu2020pre}.
In practice, these approaches have a limited scope for domains where collecting unlabeled data on a large scale is difficult due to copyright limitations or simple shortage.
In these cases, certain forms of weak supervision may compensate for the lack of data.
For example, researchers have shown that contextual metadata can be used for representation learning of biomedical images~\cite{spiegel2019metadata} or document editorial metadata for document classification~\cite{raman2021domain}.
Music is also rich in metadata, which motivates using this information to train models.


Such information can be divided into \textit{editorial metadata} used to catalog music (e.g., artist and album names, or country and year of release), and \textit{consumption metadata} describing interactions of humans (or machines) with music (e.g.,  playlists, DJ setlists, radio programs, or listening histories).
In this paper, we focus on the latter type. Using consumption metadata as a source of similarity ground truth has already been explored in the recommender-systems literature, enabling tasks such as music playlist continuation~\cite{chen2018recsys}.
Also, while editorial relations are normally one- or few-to-many (e.g., album-songs), consumption is many-to-many (e.g., playlists-songs), resulting in a more dense co-occurrence space that may favor associating more heterogeneous music.\footnote{
For example, our dataset has an average number of tracks per artist and playlist of 7.2 and 66.3, respectively.
On average, a track appears on 29.3 playlists and belongs to 1.28 artists.
}
Furthermore, the usage of consumption metadata for music representation learning has not been as extensively investigated yet~\cite{huang2020large,Ferraro2021Enriched} as the case of editorial metadata~\cite{Park2018RepresentationLO,Kim2018,Lee2019,Kim2020,huang2020large,alonso2020deep}.

\vspace{-0.7em}
\section{Method}
\label{sec:method}
\vspace{-0.7em}

We investigate methods to obtain targets from music playlist datasets to pre-train models using contrastive learning.

\vspace{-0.7em}
\subsection{Contrastive learning setup}
\begin{figure}
  \includegraphics[width=8.5cm]{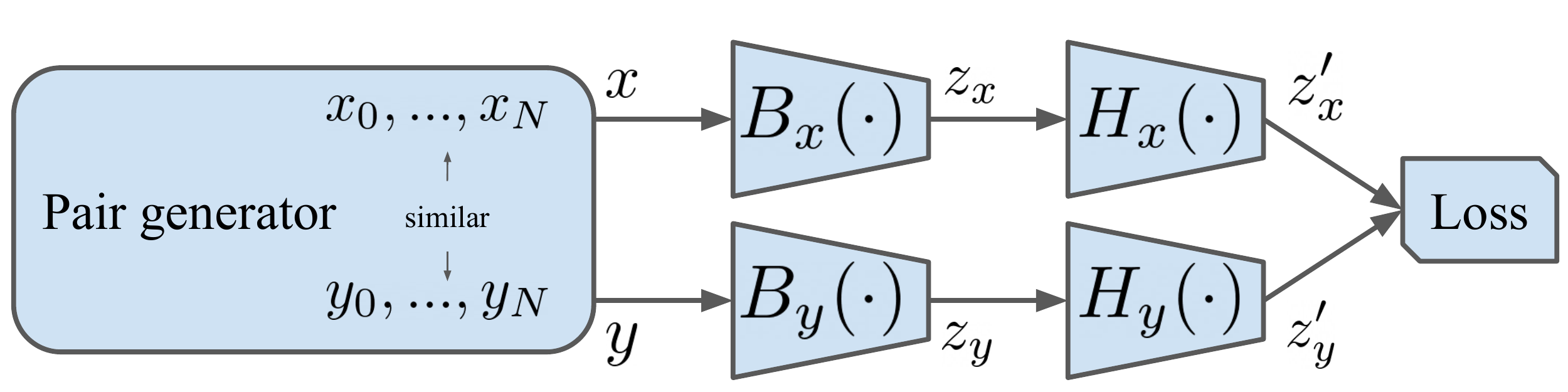}
   \caption{Illustration of our pre-training pipeline. 
   The features $x$ and $y$ from the associated pairs are input to the model $B(\cdot)$ and projector $H(\cdot)$.
   $B(\cdot)$ and $H(\cdot)$ are optimized using a contrastive loss.
   $B(\cdot)\! =\! B_x(\cdot)\! =\! B_y(\cdot)$ and $H(\cdot)\! =\! H_x(\cdot)\! =\! H_y(\cdot)$ in all the cases except for \textit{Word2Vec representation}.
  }
  \vspace{-1em}
   \label{fig:diagram}
\end{figure}

\vspace{-0.7em}
Our architecture consists of a convolutional backbone $B(\cdot)$ and a projector $H(\cdot)$ that map a mel-spectrogram input $x \in \mathbb{R}^{T \times F}$ with $T$ timestamps and $F$ frequencies into latent representations $ z \in \mathbb{R}^D$, and $z^\prime \in \mathbb{R}^{D^\prime} $ respectively.
The model is trained to bring $z^\prime_x$ close to $z^\prime_y$ while pulling it apart from samples in the same batch following SimCLR~\cite{chen2020simple}.
After pre-training, $H(\cdot)$ is discarded, and $B(\cdot)$ is used in the downstream tasks.
Our setup is depicted in Figure~\ref{fig:diagram}.

\vspace{-0.7em}
\subsection{Pair generation algorithms}
\vspace{-0.7em}
Instead of using augmentations to obtain $x$ and $y$ as done in SimCLR, we propose to use pairs originating from different tracks by exploiting playlist information.
The number of possible pairs of elements that co-occur in a playlist of size $n$ corresponds to the number of combinations without repetition ${n \choose 2}$.
This produces many pairs when considering millions of playlists, making the exhaustive usage of the pairs difficult to scale with this contrastive learning approach.
Because of this, we propose algorithms that rely on heuristics to create audio pairs that utilize a wide range of tracks while preventing track repetitions,
as well as an embedding learning-based technique to create the target pairs.

Considering a dataset of \textit{playlists} $P = \{p_0, ..., p_N\}$, and \textit{tracks} $S = \{s_{0}, ..., s_{M}\}$,
we propose the following strategies:


\begin{itemize}[noitemsep]
    
    \item \textit{Co-Occurrence}.
    This approach randomly generates pairs by producing combinations using the available tracks in each playlist $p_i$ and with each track appearing in only one pair.
    We iterate randomly through $P$ generating $ \lfloor \frac{|p_i|}{2} \rfloor$ pairs per playlist and discarding the associated tracks from the set of available tracks. This algorithm is executed at the beginning of each training epoch.
    
 
    \item \textit{Top Co-Occurrence}. This algorithm counts the number of co-occurrences of the tracks in all the playlists.
    For each track we randomly select its associated pair among its top-10 most co-occurring tracks
    while ensuring that every track appears only in one pair.
    To do so, at each epoch, we initialize a set of available tracks $A=S$.  We randomly iterate through $A$ and for a given track $s_j$ we select one of the top co-occurring tracks $s_k$ and discard $s_j$ and $s_k$ from $A$.

    \item \textit{Word2Vec representation}.
    This is a multi-modal approach in which, for a given track, we align the projection of its audio representation $z^{\prime}_x$ to the projection of its \textit{Word2Vec} embedding $z^{\prime}_y$~\cite{mikolov2013efficient}.
    We train a \textit{Word2Vec} model by considering playlists as sentences and track ids as words.
    We rely on the Continuous Bag of Words approach with a context window that includes the entire playlist~\footnote{
    We also tested a W2V sensitive to the track positions in the playlists by using smaller window sizes.
    However, this degraded the performance.}
    and a learning rate of 0.02 for 20 epochs.\footnote{We use the Gensim implementation \url{https://radimrehurek.com/gensim/models/word2vec.html}}
    In this case $B_y(\cdot)$ is the frozen pre-trained \textit{Word2Vec} model, $z_y$ is a \textit{Word2Vec} embedding, and $H_y(\cdot)$ is a different projector from $H_x(\cdot)$ featuring the same hyper-parameters and dimensions.

\end{itemize}

\vspace{-0.7em}
\section{Experiments}
\label{sec:experiments}
\vspace{-0.7em}
Our experiments are divided into two steps. First, we pre-train the proposed models in a contrastive setup.
These models are then fine-tuned and evaluated in the downstream music classification or directly evaluated in a music similarity task.




\vspace{-0.7em}
\subsection{Pre-training}
\vspace{-0.7em}


We pre-train a number of models following the different pair generation strategies.
\textit{SimCLR} is a baseline where $x$ and $y$ are alternative views of the same audio patch mixed with random patches from the batch scaled with a gain factor sampled from a $\beta(5, 2)$ distribution similar to previous work~\cite{wang2022towards}.
\textit{Artist CO} is another baseline that applies the \textit{Co-Occurrence} strategy to the artist names (i.e., considering the set of tracks by each artist as a playlist) without preventing track repetitions.

\textit{Playlist CO}, \textit{Playlist TCO}, and \textit{Playlist W2V} use the \textit{Co-Occurrence}, \textit{Top Co-Occurrence}, and \textit{Word2Vec representation} strategies respectively to generate the $x$/$y$ pairs using the playlist information.
Table~\ref{tab:training_pairs} shows the number of pairs per epoch and model. 
In \textit{SimCLR} and \textit{Playlist W2V}, the number of pairs corresponds to the number of tracks in the dataset since these methods do not associate different tracks.
In \textit{Playlist CO} and \textit{Playlist TCO}, we constrain to a single track occurrence per epoch, which results in fewer pairs per epoch.
We train the models for a fixed number of 50 epochs.
This makes the pair generation algorithms execute the same number of times, which leads to a different number of batch optimization steps for each model.


We pre-train all our models using the Million Playlist Dataset (\textit{MPD})~\cite{chen2018recsys} matched to our in-house music collection, which resulted in 1,779,072 tracks and 999,219 playlists.
$H(\cdot)$ has a single hidden layer with 128 units and a ReLu activation and $D^\prime = 128$.
We use the NT-Xent loss~\cite{chen2020simple} with a fixed $\tau$ value of 0.1 using a batch size of 384 pairs, and the Adam optimizer with $\beta_1=0.9$ and $\beta_2=0.999$.
The learning rate is increased linearly from 0 to $1e\text{-}4$ for the first 5,000 steps and then decreased following a cosine decay until the models complete 50 epochs similar to~\cite{wang2022towards}.
We train the models on 96-band, 256-timestamp ($\sim$3 seconds) mel-spectrogram patches randomly selected at each iteration from the 30-seconds excerpts available for each track.

\begin{table}[]
    \centering
    \footnotesize
    \begin{tabular}{lcc}
        \toprule
        Model & Pairs per epoch & Pair generation algorithm\\
        \midrule
        SimCLR & 1,779,072 & - \\
        Artist CO & 1,014,528$^*$ & \textit{Co-Occurrence} \\
        Playlist CO & 826,368$^*$ & \textit{Co-Occurrence} \\
        Playlist TCO & 731,520$^*$ & \textit{Top Co-Occurrence} \\
        Playlist W2V & 1,779,072 & \textit{Word2Vec representation} \\
        \bottomrule
    \end{tabular}
    \caption{Number of pairs per epoch and pair generation algorithms.
    $^*$indicates that these are different pairs on each epoch.
    }
    \vspace{-1em}
    \label{tab:training_pairs}
\end{table}

\vspace{-0.7em}
\subsection{Music classification}
\vspace{-0.7em}
Our first evaluation consists of solving multi-label music classification tasks by fine-tuning the pre-trained models.
We keep the pre-trained $B(\cdot)$ and replace $H(\cdot)$ by an MLP with 
the same hidden layer configuration
and output dimensions matching the number of classes followed by a Sigmoid activation.
We optimize $B(\cdot)$ and the new $H(\cdot)$ using Adam ($\beta_1=0.9$ and $\beta_2=0.999$) and cross-entropy loss with an L2 regularization term of $1e\text{-}5$ for a maximum of 50 epochs.
We use a cyclical triangular scheduler that varies the learning rate from $1e\text{-}5$ to $1e\text{-}4$~\cite{smith2017cyclical}.
The weights are selected from the epoch with the highest Average Precision on the validation set.
We apply early stopping after ten epochs without any improvement on this metric.
In training, we use the same random patch selection approach as in pre-training. During inference, we average the activations from non-overlapping patches.
We use 30 seconds of audio from the center of the track in validation, and the full duration available in testing.

We use the Genre, Instrument, and Mood subsets of the MTG-Jamendo Dataset~\cite{bogdanov2019jamendo}, containing 55,215, 25,135, and 18,4856 full tracks, and 87, 40, and 56 classes, respectively.
We consider the MagnaTagATune (MTAT) dataset~\cite{law2009evaluation}, with 25,860 30-seconds excerpts, and its top-50 tags using the 12:1:3 partition~\cite{van2014transfer}.
Additionally, we consider an in-house genre dataset containing 87,542 2-minutes excerpts and 72 classes, referred to as Genre Internal.
Our goal is to assess if our pre-training approaches are still beneficial when a bigger and arguably more curated collection is available. 

\vspace{-0.7em}
\subsection{Music similarity}
\vspace{-0.7em}
For the music similarity evaluation, we use the dim-sim dataset consisting of a collection of music similarity triplets produced by human raters~\cite{Lee2019MusicSimilarity}.
Each triplet was annotated by 5 to 12 people, and the official clean version of the dataset contains 879 triplets with a high inter-annotator agreement.
We extract representations $z$ for the clean subset of dim-sim using the pre-trained models without fine-tuning.
Following the common evaluation approach~\cite{Lee2019MusicSimilarity}, we measure the cosine distance between anchor/positive, and anchor/negative, and consider the triplet prediction correct if the latter is larger.
We report the prediction accuracy and the average difference between anchor/negative and anchor/positive distances.

\vspace{-0.7em}
\subsection{Architectures}
\vspace{-0.7em}
We consider two standard backbone architectures:
\begin{itemize}[noitemsep]
    \item \textbf{VGGish}~\cite{hershey2016cnn}. This is a variant of the VGG~\cite{simonyan2014very} architecture popular in the audio domain. It has 128 output dimensions.
    We consider the original model weights obtained from a classification task in a proprietary dataset as a baseline.\footnote{\url{https://github.com/tensorflow/models/tree/master/research/audioset/vggish}}
    When pre-training the architecture with our data, we use our 3-second 96-bands mel-spectrogram patches and modify the kernel of the first pooling layer from $2 \times 2$ to $4 \times 4$ to keep the number of dimensions after the convolutional layers close to the one in the original model.
    
    \item \textbf{ResNet50}~\cite{he2016deep}. We use the standard ResNet50 model considering its good performance in audio and music applications~\cite{wang2022towards}.
    We reduce the output of the last dense layer with global max- and mean-pooling and concatenate the resulting vectors, leading to an output embedding of 4,096 dimensions.
    

\end{itemize}

{%

\begin{table*}[t]
    \centering
    \footnotesize

\begin{tabular}{lP{1.05cm}P{1.05cm}c@{\hspace{0.3cm}}P{1.05cm}P{1.05cm}c@{\hspace{0.3cm}}P{1.05cm}P{1.05cm}c@{\hspace{0.3cm}}P{1.05cm}P{1.05cm}c@{\hspace{0.3cm}}P{1.05cm}P{1.05cm}}
    \toprule
        Dataset & \multicolumn{2}{c}{Genre} & & \multicolumn{2}{c}{Instrument} & & \multicolumn{2}{c}{Mood} & & \multicolumn{2}{c}{MTAT} & & \multicolumn{2}{c}{Genre Internal} \\
         & AP & ROC & & AP & ROC & & AP & ROC &&  AP & ROC & & AP & ROC \\
    \midrule
         & \multicolumn{14}{c}{VGGish}\\
\textit{VGGish FT} & 15.8$\pm$0.3 & 84.9$\pm$0.5 & & 18.1$\pm$0.6 & 74.2$\pm$1.2 & & 12.1$\pm$0.9 & 72.7$\pm$0.8 & & 44.4$\pm$0.6 & 90.6$\pm$0.1 & & - & - \\

\textit{From Scratch} & 13.4$\pm$0.2 & 82.6$\pm$0.3 & & 15.5$\pm$0.4 & 72.1$\pm$0.5 & & 9.3$\pm$0.2 & 70.6$\pm$0.4 & & 40.2$\pm$0.7 & 88.9$\pm$0.2 & & 54.3$\pm$0.3 & 96.7$\pm$0.0 \\
\textit{SimCLR} & 15.2$\pm$0.3 & 83.6$\pm$0.4 & & 16.4$\pm$0.3 & 72.1$\pm$0.5 & & 10.7$\pm$0.2 & 70.1$\pm$0.2 & & 41.1$\pm$0.6 & 88.9$\pm$0.2 & & 61.7$\pm$0.1 & 97.6$\pm$0.0 \\
\textit{Artist CO} & 17.3$\pm$0.1 & 85.6$\pm$0.1 & & 20.4$\pm$0.4 & 76.7$\pm$0.1 & & 13.9$\pm$0.2 & 74.5$\pm$0.5 & & 46.2$\pm$0.1 & 91.1$\pm$0.1 & & 68.8$\pm$0.2 & 98.3$\pm$0.1 \\
\vspace{-0.75em} \\
\textit{Playlist CO} & 17.0$\pm$0.1 & 85.4$\pm$0.2 \cellcolor[gray]{0.9} & & 20.2$\pm$0.4 \cellcolor[gray]{0.9} & 76.1$\pm$0.3 & & 13.3$\pm$0.8 \cellcolor[gray]{0.9} & 73.8$\pm$0.8 \cellcolor[gray]{0.9} & & 45.9$\pm$0.2 & 90.9$\pm$0.0 & & 67.7$\pm$0.7 & 98.2$\pm$0.1 \cellcolor[gray]{0.9} \\
\textit{Playlist TCO} & 17.5$\pm$0.1 \cellcolor[gray]{0.9} & 84.9$\pm$0.4 & & 20.5$\pm$0.3 \cellcolor[gray]{0.9} & 76.3$\pm$0.9 \cellcolor[gray]{0.9} & & 13.8$\pm$0.3 \cellcolor[gray]{0.9} & 73.7$\pm$0.7 \cellcolor[gray]{0.9} & & 45.8$\pm$0.3 & 91.0$\pm$0.1 & & 70.0$\pm$0.4 \cellcolor[gray]{0.9} & 98.4$\pm$0.0 \cellcolor[gray]{0.9} \\
\textit{Playlist W2V} & 17.3$\pm$0.2 \cellcolor[gray]{0.9} & 85.5$\pm$0.3 \cellcolor[gray]{0.9} & & 19.8$\pm$0.7 \cellcolor[gray]{0.9} & 75.3$\pm$0.2 & & 13.5$\pm$0.1 & 72.8$\pm$0.7 & & 45.3$\pm$0.6 & 90.9$\pm$0.2 \cellcolor[gray]{0.9} & & 69.8$\pm$0.1 \cellcolor[gray]{0.9} & 98.4$\pm$0.0 \cellcolor[gray]{0.9} \\

    \midrule
     & \multicolumn{14}{c}{Resnet50}\\
 
\textit{From Scratch} & 14.4$\pm$0.2 & 82.9$\pm$0.1 & & 15.6$\pm$0.5 & 71.2$\pm$0.6 & & 8.9$\pm$0.0 & 69.2$\pm$0.4 & & 40.7$\pm$0.3 & 88.8$\pm$0.1 & & 63.3$\pm$0.1 & 97.7$\pm$0.1 \\
\textit{SimCLR} & 16.3$\pm$0.2 & 84.7$\pm$0.3 \cellcolor[gray]{0.9} & & 17.4$\pm$0.1 & 73.3$\pm$0.7 & & 12.1$\pm$0.3 & 73.0$\pm$0.3 & & 43.4$\pm$0.5 & 90.1$\pm$0.2 & & 67.4$\pm$0.3 & 98.2$\pm$0.0 \\
\textit{Artist CO} & \textbf{19.0$\pm$0.1} & 85.0$\pm$0.1 & & 21.1$\pm$0.4 & 76.5$\pm$0.9 & & 14.9$\pm$0.3 & 74.8$\pm$0.7 & & 47.0$\pm$0.3 & \textbf{91.5$\pm$0.2} & & 73.4$\pm$0.2 & 98.6$\pm$0.1 \\
\vspace{-0.75em} \\
\textit{Playlist CO} & 18.7$\pm$0.7 \cellcolor[gray]{0.9} & \textbf{85.7$\pm$0.4} \cellcolor[gray]{0.9} & & \textbf{21.2$\pm$0.7} \cellcolor[gray]{0.9} & 76.7$\pm$0.9 \cellcolor[gray]{0.9} & & 14.8$\pm$0.5 \cellcolor[gray]{0.9} & 74.2$\pm$0.4 \cellcolor[gray]{0.9} & & 46.8$\pm$0.2 \cellcolor[gray]{0.9} & 91.4$\pm$0.0 \cellcolor[gray]{0.9} & & 73.4$\pm$0.1 \cellcolor[gray]{0.9} & 98.6$\pm$0.0 \cellcolor[gray]{0.9} \\
\textit{Playlist TCO} & 18.9$\pm$0.2 \cellcolor[gray]{0.9} & 85.1$\pm$0.3 \cellcolor[gray]{0.9} & & 20.4$\pm$0.7 \cellcolor[gray]{0.9} & 75.4$\pm$1.5 \cellcolor[gray]{0.9} & & 14.3$\pm$0.4 & 73.7$\pm$0.7 \cellcolor[gray]{0.9} & & \textbf{47.0$\pm$0.2} \cellcolor[gray]{0.9} & 91.3$\pm$0.2 \cellcolor[gray]{0.9} & & 72.8$\pm$0.2 & 98.6$\pm$0.0 \cellcolor[gray]{0.9} \\
\textit{Playlist W2V} & \textbf{19.0$\pm$0.1} \cellcolor[gray]{0.9} & 85.4$\pm$0.3 \cellcolor[gray]{0.9} & & 20.7$\pm$0.4 \cellcolor[gray]{0.9} & \textbf{77.1$\pm$0.4} \cellcolor[gray]{0.9} & & \textbf{15.0$\pm$0.1} \cellcolor[gray]{0.9} & \textbf{75.1$\pm$0.4} \cellcolor[gray]{0.9} & & 46.7$\pm$0.4 \cellcolor[gray]{0.9} & 91.2$\pm$0.2 \cellcolor[gray]{0.9} & & \textbf{74.1$\pm$0.2} \cellcolor[gray]{0.9} & \textbf{98.7$\pm$0.1} \cellcolor[gray]{0.9} \\

    \bottomrule
    \end{tabular}
    \caption{Metrics in the music classification datasets expressed in macro ROC-AUC and Average Precision.
    For each architecture, we present the baselines on top and the proposed models below.
    Metrics statistically equivalent or higher than \textit{Artist CO}
    according to a one-sided t-test ($p\text{-}value=0.005$)
    are marked in light grey.
    The highest metric per dataset is marked in bold.
    }
    \vspace{-1em}
    \label{tab:classification}
\end{table*}
}


\begin{table}[]
    \centering
    \footnotesize
    \begin{tabular}{lllll}
    \toprule
        Model & VGGish & Resnet50 & VGGish & Resnet50 \\
        & \multicolumn{2}{c}{Accuracy} & \multicolumn{2}{c}{Average difference} \\       
    \midrule
        \textit{SimCLR} & 0.699 & 0.672 & 0.007 & 0.009 \\
        \textit{Artist CO} & 0.819 & 0.838 & 0.043 &  0.039 \\
        \textit{Playlist CO} & \textbf{0.852} &  \textbf{0.845} & \textbf{0.077} & \textbf{0.064} \\
        \textit{Playlist TCO} & 0.793 & 0.813 & 0.052 &  0.046 \\
        \textit{Playlist W2V} & 0.831 & 0.818 & 0.067 & 0.041  \\
    \bottomrule
    \end{tabular}
    \caption{Music similarity accuracy and the average difference between anchor/negative and anchor/positive.}
    \vspace{-1em}
    \label{tab:similarity}
\end{table}


\vspace{-0.7em}
\section{Results and discussion}
\label{sec:results}
\vspace{-0.7em}

Table~\ref{tab:classification} shows the macro ROC-AUC and Average Precision\footnote{Average Precision is also referred to as the area under the precision-recall curve (PR-AUC) in the literature.} metrics for all the datasets and models as the average $\pm$ the standard deviation of three runs.
Our baselines consist of fine-tuning the original VGGish~\cite{hershey2016cnn} (\textit{VGGish FT}), randomly initialized models (\textit{From Scratch}), \textit{SimCLR}, and \textit{Artist CO}.




Firstly, we note that the contrastive approaches based on artist and playlist metadata
always achieve better performance than the \textit{VGGish FT}, \textit{From Scratch}, and \textit{SimCLR} baselines, which aligns with previous works indicating the benefits of metadata-based supervision~\cite{Park2018RepresentationLO,Kim2018,Lee2019,alonso2022music}.
The models based on playlist information achieve equivalent or superior performance to those based on \textit{Artist CO} on most datasets and metrics, and \textit{Playlist W2V} with the ResNet50 architecture achieves the best performance in at least one metric for the Genre, Instrument, Mood, and Genre Internal datasets.




Table~\ref{tab:similarity} contains the results of the music similarity evaluation.
We observe that models based on metadata show a stronger correlation with human  similarity perception than the baseline SimCLR approach.
While \textit{Playlist CO} achieves the best metrics with both architectures, \textit{Playlist TCO} and \textit{Playlist W2V} did not improve the performance as in the classification tasks.
We hypothesize that \textit{Top Co-Occurrence} and \textit{Word2Vec representation} reduce the diversity of the positive pairs, which may augment the discriminative capabilities of the latent space at the cost of becoming weaker for similarity.

Finally, these results may depend on the nature and sparsity of the available playlists. 
In our study, we relied on MPD, which contains a curated subset of Spotify playlists filtered by quality and enriched with additional tracks.
The playlists were created by US users only between 2010 and 2017 and are not expected to be representative of the overall distribution of Spotify playlists.
However, MPD represents a small fraction of more than 4 billion playlists on Spotify, which motivates further research on playlist-based pre-training.



\vspace{-0.8em}
\section{Conclusions}
\label{sec:conclusion}
\vspace{-0.7em}
In this work, we show that employing contrastive learning for pre-training neural networks with playlist information is valuable for music classification.
While previous works focused on editorial metadata, such as the artist name, we found that superior performance can be achieved with consumption metadata consisting of playlist information
by relying on track representations obtained from a Word2Vec model trained on the playlist sequences.
Also, the representations learned using simple playlist co-occurrences perform significantly better than an unsupervised approach (SimCLR) or than using artist co-occurrences for music similarity.
Future work includes validating our approaches with more sources of consumption metadata (e.g., radio programs or listening histories) considering learning them in multi-task scenarios. 

\vspace{-1em}
\section{Acknowledgements}
\vspace{-0.8em}
\footnotesize
This research was partially funded by Musical AI - PID2019-111403GB-I00/AEI/10.13039/501100011033 of the Spanish Ministerio de Ciencia, Innovación y Universidades.

\bibliographystyle{IEEEbib}
{
\footnotesize
\bibliography{refs}
}

\end{document}